\documentclass[a4paper,12pt]{article}
\usepackage{amsmath}
\usepackage{latexsym}
\begin{document}
\begin{titlepage}
\begin{flushright}
gr-qc/0009074\\
\textbf{UA/NPPS-7}
\end{flushright}

\begin{centering}
\vspace*{1.cm}

{\large{\bf Conditional Symmetries, the True Degree of Freedom}}\\
{\large{\bf and G.C.T. Invariant Wave functions}}\\
{\large{\bf for the general Bianchi Type II Vacuum Cosmology}}

\vspace*{1.cm} {\bf T. Christodoulakis$^{\dag}$, G.O.
Papadopoulos$^{\ddag}$}

\vspace{.7cm} {\it University of Athens, Physics Department
Nuclear \& Particle Physics Section\\
GR--15771  Athens, Greece}\\
\end{centering}

\vspace{2.cm} \abstract{The quantization of the most general
Bianchi Type II geometry --with all six scale factors, as well as
the lapse function and the shift vector, present-- is considered.
In an earlier work, a first reduction of the initial
6-dimensional configuration space, to a 4-dimensional one, has
been achieved by the usage of the information furnished by the
quantum form of the linear constraints. Further reduction of the
space in which the wave function --obeying the Wheeler-DeWitt
equation-- lives, is accomplished by unrevealling the extra
symmetries of the Hamiltonian. These symmetries appear in the
form of --linear in momenta-- first integrals of motion. Most of
these symmetries, correspond to G.C.T.s through the action of the
automorphism group. Thus, a G.C.T. invariant wave function is
found, which depends on the only true degree of freedom, i.e. the
unique curvature invariant, characterizing the hypersurfaces
$t=const$.}

\vspace{2.2cm} \noindent
\rule[0.cm]{11.6cm}{.009cm} \\
\vspace{.3cm} e-mail: \, $\dag$ tchris@cc.uoa.gr, $\ddag$
gpapado@cc.uoa.gr
\end{titlepage}
\baselineskip=19pt

\newpage

\section{}
As it is well known, the quantum cosmology approximation,
consists in freezing out all but a finite number of degrees of
freedom of the gravitational field, and quantize the rest. This
is done by imposing spatial homogeneity. Thus our --in principle--
dynamical degrees of freedom, are the scale factors
$\gamma_{\alpha\beta}(t)$, the lapse function $N(t)$ and the
shift vector $N^{a}(t)$, of some Bianchi Type geometry.

The general Bianchi Type II cosmology, had been treated in
\cite{1}, where a wave function --in terms of 4 combinations of
the 6 $\gamma_{\alpha\beta}(t)$-- had been presented. According
to the hints given in the discussion of \cite{1}, this reduction
ought not to be the final one, since there were gauge degrees of
freedom left.

In this sort communication, we present the desired final
reduction and the only true degree of freedom is revealed as an
argument of the wave function.

\section{}
In \cite{1}, we had considered the quantization of an action corresponding to the most general
Bianchi Type II cosmology, i.e. an action giving Einstein's field equations, derived from the line
element:
\begin{equation}
ds^{2}=(N^{2}(t)-N_{a}(t)N^{a}(t))dt^{2}+2N_{a}(t)\sigma^{a}_{i}(x)dx^{i}dt+
\gamma_{\alpha\beta}(t)\sigma^{\alpha}_{i}(x)\sigma^{\beta}_{j}(x)dx^{i}dx^{j}
\end{equation}
with:
\begin{equation}
\begin{array}{l}
  \sigma^{a}(x)=\sigma^{\alpha}_{i}(x)dx^{i}\\
  \sigma^{1}(x)=dx^{2}-x^{1}dx^{3} \\
  \sigma^{2}(x)=dx^{3} \\
  \sigma^{3}(x)=dx^{1} \\
  d\sigma^{a}(x)=\frac{1}{2}C^{a}_{\beta\gamma}\sigma^{\beta}\wedge\sigma^{\gamma}\\
  C^{1}_{23}=-C^{1}_{32}=1
\end{array}
\end{equation}
see \cite{2}.

As is well known \cite{3}, the Hamiltonian is
$H=N(t)H_{0}+N^{a}(t)H_{a}$ where:
\begin{equation}
H_{0}=\frac{1}{2}L_{\alpha\beta\mu\nu}\pi^{\alpha\beta}\pi^{\mu\nu}+\gamma
R
\end{equation}
is the quadratic constraint with:
\begin{displaymath}
\begin{array}{l}
  L_{\alpha\beta\mu\nu}=\gamma_{\alpha\mu}\gamma_{\beta\nu}+\gamma_{\alpha\nu}\gamma_{\beta\mu}-
\gamma_{\alpha\beta}\gamma_{\mu\nu} \\
  R=C^{\beta}_{\lambda\mu}C^{\alpha}_{\theta\tau}\gamma_{\alpha\beta}\gamma^{\theta\lambda}
\gamma^{\tau\mu}+2C^{\alpha}_{\beta\delta}C^{\delta}_{\nu\alpha}\gamma^{\beta\nu}+
4C^{\mu}_{\mu\nu}C^{\beta}_{\beta\lambda}\gamma^{\nu\lambda}=
C^{\alpha}_{\mu\kappa}C^{\beta}_{\nu\lambda}\gamma_{\alpha\beta}\gamma^{\mu\nu}\gamma^{\kappa\lambda}
\end{array}
\end{displaymath}
$\gamma$ being the determinant of $\gamma_{\alpha\beta}$ (the last
equality holding only for the Type II case), and:
\begin{equation}
H_{a}=C^{\mu}_{a\rho}\gamma_{\beta\mu}\pi^{\beta\rho}
\end{equation}
are the linear constraints.

The quantities $H_{0}$, $H_{a}$, are weakly vanishing \cite{4},
i.e. $H_{0}\approx 0$, $H_{a}\approx 0$. For all class A Bianchi
Types ($C^{\alpha}_{\alpha\beta}=0$), it can be seen to obey the
following first-class algebra:
\begin{equation}
\begin{array}{l}
  \{H_{0}, H_{0}\}=0 \\
  \{H_{0}, H_{a}\}=0 \\
  \{H_{a}, H_{\beta}\}=-\frac{1}{2}C^{\gamma}_{\alpha\beta}H_{\gamma}
\end{array}
\end{equation}
which ensures their preservation in time i.e. $\dot{H}_{0}\approx
0$, $\dot{H}_{a}\approx 0$ and establishes the consistency of the
action.

If we follow Dirac's general proposal \cite{4} for quantazing
this action, we have to turn $H_{0}$, $H_{a}$, into operators
annihilating the wave function $\Psi$.

In the Schr\"{o}dinger representation:
\begin{equation}
\begin{array}{l}
  \gamma_{\alpha\beta}\rightarrow
\widehat{\gamma}_{\alpha\beta}=\gamma_{\alpha\beta} \\
  \pi^{\alpha\beta}\rightarrow
\widehat{\pi}^{\alpha\beta}=-i\frac{\partial}{\partial\gamma_{\alpha\beta}}
\end{array}
\end{equation}
satisfying the basic Canonical Commutation Relation (CCR)
--corresponding to the classical ones:
\begin{equation}
[\widehat{\gamma}_{\alpha\beta},
\widehat{\pi}^{\mu\nu}]=-i\delta^{\mu\nu}_{\alpha\beta}=\frac{-i}{2}
(\delta^{\mu}_{\alpha}\delta^{\nu}_{\beta}+\delta^{\mu}_{\beta}\delta^{\nu}_{\alpha})
\end{equation}

The quantum version of the 2 independent linear constraints has
been used to reduce, via the method of characteristics \cite{5},
the dimension of the initial configuration space from 6
($\gamma_{\alpha\beta}$) to 4 (combinations of
$\gamma_{\alpha\beta}$), i.e.
$\Psi=\Psi(q,\gamma,\gamma^{2}_{12}-\gamma_{11}\gamma_{22},\gamma_{12}\gamma_{13}-
\gamma_{11}\gamma_{23})$, where
$q=C^{\alpha}_{\mu\kappa}C^{\beta}_{\nu\lambda}\gamma_{\alpha\beta}\gamma^{\mu\nu}\gamma^{\kappa\lambda}$.

According to K\v{u}char's and Hajicek's \cite{6} prescription,
the ``kinetic'' part of $H_{0}$ is to be realized as the
conformal Laplacian, corresponding to the reduced metric:
\begin{equation}
L_{\alpha\beta\mu\nu}\frac{\partial x^{i} }{\partial
\gamma_{\alpha\beta} }\frac{\partial x^{j}}{\partial
\gamma_{\mu\nu}}=g^{ij}
\end{equation}
where $x^{i}$, $i=1,2,3,4$, are the arguments of $\Psi$. The
solutions had been presented in \cite{1}. Note that the
first-class algebra satisfied by $H_{0}$, $H_{a}$, ensures that
indeed, all components of $g^{ij}$ are functions of the $x^{i}$.
The signature of the $g^{ij}$, is $(+, +, -, -)$ signaling the
existence of gauge degrees of freedom among the $x^{i}$'s.

Indeed, one can prove \cite{7} that the only gauge invariant
quantity which, uniquely and irreducibly, characterizes a
3-dimensional geometry admitting the Type II symmetry group, is:
\begin{equation}
q=C^{\alpha}_{\mu\kappa}C^{\beta}_{\nu\lambda}\gamma_{\alpha\beta}\gamma^{\mu\nu}\gamma^{\kappa\lambda}
\end{equation}
An outline of the proof, is as follows:\\
Let two exads $\gamma^{(1)}_{\alpha\beta}$ and
$\gamma^{(2)}_{\alpha\beta}$ be given, such that their
corresponding $q$'s, are equal. Then \cite{7} there exists an
automorphism matrix $\Lambda$ (i.e. satisfying
$C^{a}_{\mu\nu}\Lambda^{\kappa}_{a}=C^{\kappa}_{\rho\sigma}\Lambda^{\rho}_{\mu}\Lambda^{\sigma}_{\nu}$)
connecting them, i.e.
$\gamma^{(1)}_{\alpha\beta}=\Lambda^{\mu}_{\alpha}\gamma^{(2)}_{\mu\nu}\Lambda^{\nu}_{\beta}$.
But as it had been shown in the appendix of \cite{8}, this kind of
changes on $\gamma_{\alpha\beta}$, can be seen to be induced by
spatial diffeomorphisms. Thus, 3-dimensional Type II geometry, is
uniquely characterized by some value of $q$.

Although for full pure gravity, K\v{u}char \cite{9} has shown that
there are not other first-class functions, homogeneous and linear
in $\pi^{\alpha\beta}$, except $H_{a}$, imposing the extra
symmetries (Type II), allows for such quantities to exist --as
it will be shown. We are therefore, naturally led to seek the
generators of these extra symmetries --which are expected to chop
off $x^{2}$, $x^{3}$, $x^{4}$. Such quantities are, generally,
called in the literature ``Conditional Symmetries''.

The automorphism group for Type II, is described by the following
6 generators --in matrix notation and collective form:
\begin{equation}
\lambda^{a}_{(I)\beta}=\left(
\begin{array}{ccc}
  \kappa+\mu & x & y \\
  0 & \kappa & \rho \\
  0 & \sigma & \mu
\end{array}\right)
\end{equation}
with the property:
\begin{equation}
C^{a}_{\mu\nu}\lambda^{\kappa}_{a}=C^{\kappa}_{\mu\sigma}\lambda^{\sigma}_{\nu}+C^{\kappa}_{\sigma\nu}\lambda^{\sigma}_{\mu}
\end{equation}
From these matrices, we can construct the linear --in momenta--
quantities:
\begin{equation}
A_{(I)}=\lambda^{a}_{(I)\beta}\gamma_{\alpha\rho}\pi^{\rho\beta}
\end{equation}
Two of these, are the $H_{a}$,'s since $C^{a}_{(\rho)\beta}$
correspond to the inner automorphism subgroup --designated by the
x and y parameters, in $\lambda^{a}_{(I)\beta}$. The rest of
them, are the generators of the outer automorphisms and are
described by the matrices:
\begin{equation}
\varepsilon^{a}_{(I)\beta}=\left(\begin{array}{ccc}
  \kappa+\mu & 0 & 0 \\
  0 & \kappa & \rho \\
  0 & \sigma & \mu
\end{array}\right)
\end{equation}
The corresponding --linear in momenta-- quantities, are:
\begin{equation}
E_{(I)}=\varepsilon^{a}_{(I)\beta}\gamma_{\alpha\rho}\pi^{\rho\beta}
\end{equation}
The algebra of these --seen as functions on the phase space,
spanned by $\gamma_{\alpha\beta}$ and $\pi^{\mu\nu}$--, is:
\begin{equation}
\begin{array}{l}
  \{E_{I}, E_{J}\}=\widetilde{C}^{K}_{IJ}E_{K} \\
  \{E_{I}, H_{a}\}=-\frac{1}{2}\lambda^{\beta}_{a}H_{\beta} \\
  \{E_{I}, H_{0}\}=-2(\kappa+\mu)\gamma R
\end{array}
\end{equation}
From the last of (15), we conclude that the subgroup of $E_{I}$'s
with the property $\kappa+\mu=0$, i.e. the traceless generators,
are first-class quantities; their time derivative vanishes. So
let:
\begin{equation}
\widetilde{E}_{I}=\{E_{I}:~\kappa+\mu=0\}
\end{equation}
Then, the previous statement translates into the form:
\begin{equation}
\dot{\widetilde{E}_{I}}=0 \Rightarrow \widetilde{E}_{I}=c_{I}
\end{equation}
the $c_{I}$'s being arbitrary constants.

Now, these are --in principle-- integrals of motion. Since, as we
have earlier seen, $\widetilde{E}_{I}$'s along with $H_{a}$'s,
generate automorphisms, it is natural to promote the integrals of
motion (17), to symmetries --by setting the $c_{I}$'s zero. The
action of the quantum version of these $\widetilde{E}_{I}$'s on
$\Psi$, is taken to be \cite{6}:
\begin{equation}
\begin{array}{l}
  \widehat{\widetilde{E}}_{I}\Psi=\varepsilon^{a}_{(I)\beta}\gamma_{\alpha\rho}\frac{\partial\Psi}{\gamma_{\beta\rho}}=0 \\
  \varepsilon^{a}_{(I)a}=0
\end{array} \Bigg\}\Rightarrow \Psi=\Psi(q,\gamma)
\end{equation}

The Wheeler-DeWitt equation now, reads:
\begin{equation}
5q^{2}\frac{\partial^{2} \Psi}{\partial
q^{2}}-3\gamma^{2}\frac{\partial^{2} \Psi}{\partial
\gamma^{2}}+2q\gamma\frac{\partial^{2}\Psi}{\partial
\gamma\partial q}+5q\frac{\partial \Psi}{\partial
q}-3\gamma\frac{\partial \Psi}{\partial \gamma}-2q\gamma\Psi=0
\end{equation}
\textit{Note that:
\begin{displaymath}
\nabla^{2}_{c}=\nabla^{2}+\frac{(d-2)}{4(d-1)}R=\nabla^{2}
\end{displaymath}
since we have a 2-dimensional, flat space, with contravariant
metric:
\begin{equation}
g^{ij}=\left(\begin{array}{cc}
  5q^{2} & q\gamma \\
  q\gamma & -3\gamma^{2}
\end{array}\right)
\end{equation}
which is Lorentzian}. This equation, can be easily solved by
separation of variables; transforming to new coordinates
$u=q\gamma^{3}$ and $v=q\gamma$, we get the 2 independent
equations:
\begin{equation}
\begin{array}{l}
  16u^{2}A''(u)+4uA'(u)-cA(u)=0 \\
  B''(v)+\frac{1}{2v}B'(v)-(\frac{1}{2v}+\frac{c}{4v^{2}})B(v)=0
\end{array}
\end{equation}
where c, is the separation constant. Equation (19), is of
hyperbolic type and the resulting wave function will still not be
square integrable --under any measure. Besides that, the
tracefull generators of the outer automorphisms, are left
inactive --due to the non vanishing CCR with $H_{0}$.

These two facts, lead us to deduce that there must still exist a
gauge symmetry, corresponding to some --would be, linear in
momenta-- first-class quantity. Our starting point in the pursuit
of this, is the third of (15). It is clear that we need another
quantity --also linear in momenta-- with an analogous property;
the trace of $\pi^{\mu\nu}$, is such an object. We thus define
the following quantity:
\begin{equation}
T=E_{I}-(\kappa+\mu)\gamma_{\alpha\beta}\pi^{\alpha\beta}
\end{equation}
in the phase space --spanned by $\gamma_{\alpha\beta}$ and
$\pi^{\mu\nu}$. It holds that:
\begin{equation}
\begin{array}{l}
  \{T, H_{0}\}=0 \\
  \{T, H_{a}\}=0 \\
  \{T, E_{I}\}=0
\end{array}
\end{equation}
because of:
\begin{equation}
\begin{array}{l}
  \{E_{I}, \gamma\}=-2(\kappa+\mu)\gamma \\
  \{E_{I}, q\}=0 \\
  \gamma_{\alpha\beta}\{\pi^{\alpha\beta}, q\}=q \\
  \gamma_{\alpha\beta}\{\pi^{\alpha\beta}, \gamma\}=-3\gamma
\end{array}
\end{equation}

Again --as for $\widetilde{E}_{I}$'s--, we see that since $T$, is
first-class, we have that:
\begin{equation}
\dot{T}=0 \Rightarrow T=const=c_{T}
\end{equation}
another integral of motion. We therefore see, that $T$ has all
the necessary properties to be used in lieu of the tracefull
generator, as a symmetry requirement on $\Psi$. In order to do
that, we ought to set $c_{T}$ zero --exactly as we did with the
$c_{I}$'s, corresponding to $\widetilde{E}_{I}$'s. The quantum
version of $T$, is taken to be:
\begin{equation}
\widehat{T}=\lambda^{\alpha}_{\beta}\gamma_{\alpha\rho}\frac{\partial}{\partial
\gamma_{\beta\rho}}-(\kappa+\mu)\gamma_{\alpha\beta}\frac{\partial}{\partial
\gamma_{\alpha\beta}}
\end{equation}
Following, Dirac's theory, we require:
\begin{equation}
\widehat{T}\Psi=\lambda^{\alpha}_{\beta}\gamma_{\alpha\rho}\frac{\partial
\Psi}{\partial
\gamma_{\beta\rho}}-(\kappa+\mu)\gamma_{\alpha\beta}\frac{\partial
\Psi}{\partial
\gamma_{\alpha\beta}}=(\kappa+\mu)(q\frac{\partial\Psi}{\partial
q}-\gamma\frac{\partial\Psi}{\partial \gamma})=0
\end{equation}
Equation (27), implies that $\Psi(q,\gamma)=\Psi(q\gamma)$ and
thus equation (19), finally, reduces to:
\begin{equation}
4w^{2}\Psi''(w)+4w\Psi'(w)-2w\Psi=0
\end{equation}
where, for simplicity, $w\doteq q\gamma$. The solution to this
equation, is:
\begin{equation}
\Psi=c_{1}I_{0}(\sqrt{2q\gamma})+c_{2}K_{0}(\sqrt{2q\gamma})
\end{equation}
where $I_{0}$ is the modified Bessel function, of the first kind,
and $K_{0}$ is the modified Bessel function, of the second kind,
both with zero argument.

At first sight, it seems that although we have apparently
exhausted the symmetries of the system, we have not yet been able
to obtain a wave function on the space of the 3-geometries, since
$\Psi$ depends on $q\gamma$ and not on $q$ only. On the other
hand, the fact that we have achieved a reduction to one degree of
freedom, must somehow imply that the wave function found must be
a function of the geometry. This puzzle finds its resolution as
follows. Consider the quantity:
\begin{equation}
\Omega=-2\gamma_{\rho\sigma}\pi^{\rho\sigma}+{
\frac{2C^{a}_{\mu\kappa}C^{\beta}_{\nu\lambda}\gamma^{\kappa\lambda}\gamma^{\mu\nu}
\gamma_{\alpha\rho}\gamma_{\beta\sigma}-4C^{\alpha}_{\mu\rho}C^{\beta}_{\nu\sigma}\gamma_{\alpha\beta}\gamma^{\mu\nu}}{q}}\pi^{\rho\sigma}
\end{equation}
This can also be seen to be first-class, i.e.
\begin{equation}
\dot{\Omega}=0 \Rightarrow \Omega=const=c_{\Omega}
\end{equation}
Moreover, is a linear combination of $T$, $\widetilde{E}_{I}$'s,
and $H_{a}$'s, and thus $c_{\Omega}=0$. Now it can be verified that
$\Omega$, is nothing but:
\begin{equation}
\frac{1}{N(t)}(\frac{\dot{\gamma}}{\gamma}+\frac{1}{3}\frac{\dot{q}}{q})
\end{equation}
So:
\begin{equation}
\gamma q^{1/3}=\vartheta=constant
\end{equation}
Without any loss of generality, and since $\vartheta$ is not an
essential constant of the classical system (see \cite{10} and
reference [18] therein), we set $\vartheta=1$. Therefore:
\begin{equation}
\Psi=c_{1}I_{0}(\sqrt{2}q^{1/3})+c_{2}K_{0}(\sqrt{2}q^{1/3})
\end{equation}
where $I_{0}$ is the modified Bessel function, of the first kind,
and $K_{0}$ is the modified Bessel function, of the second kind,
both with zero argument.

As for the measure, it is commonly accepted that, there is not a
unique solution. A natural choice, is to adopt the measure that
makes the operator in (28), hermitian --that is:
\begin{equation}
\mu(q)\propto q^{-1}
\end{equation}
It is easy to find combinations of $c_{1}$ and $c_{2}$ so that
the probability $\mu(q)|\Psi|^{2}$, be defined.

\section{}
In this work, we were able to express the wave function for a
Bianchi Type II Vacuum cosmology, in terms of the only true
degree of freedom, i.e. the only curvature invariant ($q$) of the
3-geometry, under discussion. This was done by imposing the
quantum versions of the first-class quantities
$\widetilde{E}_{I}$'s, $T$ and $\Omega$, as conditions on the wave
function $\Psi$. A crucial point in this procedure, was the
setting of the numerical values of $\widetilde{E}_{I}$'s, $T$ and
$\Omega$, equal to zero. The arguments in favor of this action,
as far as $\widetilde{E}_{I}$'s are concerned, are overwhelming
since they are generators of automorphisms which, in turn, are
induced by spatial G.C.T.s. Since $T$, is to replace the
tracefull outer automorphism generator, it is mandatory to also
set it zero. Once this has been done, $\Omega$ --being a linear
function of $H_{a}$'s, $\widetilde{E}_{I}$'s and $T$-- is inevitably
zero.

An additional argument for setting $T$ and $\Omega$ zero, can be
based on Dirac's theory: any first-class constraint is generator
of a contact transformation. Moreover, any first-class quantity,
is strongly equal to a linear combination of the constraints of
the system. So, $T$ and $\Omega$, could be considered as
generators of contact transformations i.e. as generators of
covariance transformations, analogous to the automorphisms
generated by ($H_{a}, \widetilde{E}_{I}$).

Note that putting the constant associated with $\Omega$, equal to
zero, amounts in restricting to a subset of the classical
solutions, since $c_{\Omega}$, is one of the two essential
constants of Taub's solution. One could keep that constant, at
the expense of arriving at a wave function with explicit time
dependence, since then:
\begin{displaymath}
\gamma=q^{-1/3}Exp[\int{c_{\Omega}N(t)}dt]
\end{displaymath}
We however, consider more appropriate to set that constant zero,
thus arriving at a $\Psi$ depending on $q$ only, and decree its
applicability to the entire space of the classical solutions.
Anyway this is not such a blunder, since $\Psi$ is to give weight
to all states, --being classical ones, or not.

So in conclusion, we see that not only the true degree of freedom
is isolated but also the time problem has been solved --in the
sense that a square integrable wave function $\Psi$ is found.
This is accomplished by unrevealling all the --hidden-- gauge
symmetries of the system. That wave function, is well defined on
any spatially homogeneous 3-geometry.

A similar situation holds for Class A Types VI and VII; an object
analogous to $T$ also exists, and thus a reduction to the
variables $\gamma q^{1}$, $\gamma q^{2}$ is possible with the
help of this $T$ and the 3 independent $H_{a}$'s ($q^{1}=q$,
$q^{2}=C^{\alpha}_{\beta\mu}C^{\beta}_{\alpha\nu}\gamma^{\mu\nu}$,
are the 2 independent curvature invariants, for these homogeneous
3-geometries). However, the passage to a G.C.T. invariant
wave function i.e. $\Psi=\Psi(q^{1},~q^{2})$, requires the knowledge
of a first integral analogous to $\Omega$, a thing that we luck
--at present. The situation concerning Types VIII and IX, is more
difficult: there are no outer automorphisms, and consequently no
object analogous to $T$, exists. The $H_{a}$'s suffice to reduce
the configuration space to $q^{1}$, $q^{2}$ and
$q^{3}=Det[m]/\sqrt{\gamma}$, now needed to specify the
3-geometry. The reduced supermetric is still Lorentzian, leading
to a hyperbolic Wheeler-DeWitt equation. This fact is generally
considered as a drawback, since it prohibits the ensuing wave
function from being square integrable. Within the spirit of this
work however, it can be taken as a positive sing; an object
analogous to $\Omega$ may exist. In this case we would be able to
reduce to a spacelike surface. We hope to return soon, with
concrete results on these issues.

\end{document}